\documentclass[preprint]{elsarticle}

\biboptions{sort&compress}
\usepackage{lineno,hyperref}
\usepackage{caption}
\usepackage{graphicx}
\usepackage{subfigure}
\graphicspath{{figures/}}
\modulolinenumbers[5]

\usepackage[namelimits]{amsmath}
\usepackage{amssymb}

\usepackage{booktabs}
\usepackage{multirow}
\usepackage{array}

\journal{}

\begin{document}

\begin{frontmatter}

\title{Industrial computed tomography based intelligent non-destructive testing method for power capacitor}

\author[mymainaddress]{Zhenxing Cheng\corref{mycorrespondingauthor}}
\cortext[mycorrespondingauthor]{Corresponding author}
\ead{zxcheng@hnu.edu.cn}

\author[mymainaddress]{Peng Wang}
\author[mymainaddress]{Yue Liu}
\author[mymainaddress]{Wei Qin}
\author[mysecondaryaddress]{Zidi Tang}

\address[mymainaddress]{ZhuZhou CRRC Times Electric Co., Ltd. Zhuzhou, 412001, China}
\address[mysecondaryaddress]{University of Sheffield, Sheffield, UK}

\begin{abstract}

Power capacitor device is a widely used reactive power compensation equipment in power transmission and distribution system which can easily have internal fault and therefore affects the safe operation of the power system. An intelligent non-destructive testing (I-NDT) method based on ICT is proposed to test the quality of power capacitors automatically in this study. The internal structure of power capacitors would be scanned by the ICT device and then defects could be recognized by the SSD algorithm. Moreover, the data data augmentation algorithm is used to extend the image set to improve the stability and accuracy of the trained SSD model.
	
\end{abstract}

\begin{keyword}
Power capacitor \sep Object detection \sep Single shot multibox detector \sep Industrial computed tomography \sep Non-destructive test
\end{keyword}

\end{frontmatter}


\section{Introduction}

Electrical equipment which uses power capacitors as the core component, such as reactive power compensation
device and harmonic filter device, has advantages of low cost, flexible configuration, and high quality in power system. \cite{feng2014research,wang2018application} However, the long-running power capacitor is prone to cause internal faults because of aging, overload (overheating) and so on, which affects the safe operation of the electric grid. Nowadays, power capacitor device usually uses the traditional, periodic planned testing or the current differential and voltage differential method to protect device, lacking effective means of monitoring abnormal conditions of real-time online equipment during operation, which is hardly to discover the internal failure of the device \cite{khani2010new,carvalho2016virtual,cheng2016design}. Therefore, how to discover the internal failure is an essential process to promise the quality and reliability of power capacitors. The development of industrial Computed Tomography (ICT) makes this process possible.

Computed tomography (CT) is a well-established technology in medical diagnostics\cite{hsieh2003computed}. For a few decades now, dedicated CT systems have also been in use for dimensional measurements in industry\cite{bartscher2007enhancement}. The development of ICT imaging technology has provided great convenience for non-destructive Testing (NDT) of industrial equipment\cite{Thompson_2016,sun2012overview}. Typical areas of use for CT in industry are in the detection of flaws such as voids and cracks, and the particle analysis in materials. In metrology, CT allows measurements of the external as well as the internal geometry of complex parts. So far, CT metrology is the only technology able to measure as well the inner as the outer geometry of a component without cutting or destroying it. As such, it is the only technology for industrial quality control of work-pieces having non-accessible internal features (e.g. components produced by additive manufacturing) or multi-material components (e.g. two-component injection molded plastic parts or plastic parts with metallic inserts). CT can be considered as a third revolution in coordinate metrology, following the introduction of tactile 3D coordinate measuring machines (CMMs) in the seventies and that of optical 3D scanners in the eighties \cite{de2014industrial}.

Usually, the quality of a work-piece is decided by technicians from ICT images. The defects in the image are usually recognized by human eyes and then the testing quality level can be determined by technicians. However, those methods are inconvenient when meeting a large number of testing tasks. Moreover, the testing level would be confused due to the effect of human emotions, visual fatigue and other factors. Therefore, it is important to propose a method for objective testing, such as an object detection based non-destructive testing method in which object detection algorithms are used to recognize defects from those ICT images. Single Shot MultiBox Detector algorithm proposed by Liu et al.\cite{liu2016ssd} is a single deep neural network, combined with anchor points, RPN and multi-scale representation\cite{zhang2021comprehensive}, which is used to detect objects from images. Single Shot MultiBox Detector has been applied to all walk of life, such as  face recognition\cite{nagrath2021ssdmnv2},  automatic drive\cite{arora2019real}, rail transit\cite{chen2018fast}, video surveillance\cite{huang2019improved,kumar2020object,magalhaes2021evaluating} and so on.

The intelligent non-destructive testing technology based on ICT is proposed to test and rate the quality of parts automatically in this project. The corresponding feature images should be extracted from the ICT images by the specific image processing technology, and then the defects in the extracted image could be intelligently recognized by the deep convolutional neural network (DCNN). Finally, the rank of the quality should be determined by the algorithms according to the analysis result of DCNN. 

The purpose of this study is to find an effective method to achieve internal defect diagnosis of power capacitors. In this study, an intelligent non-destructive testing (I-NDT) method based on ICT is proposed to test the quality of workpieces automatically. The corresponding feature images of power capacitors should be extracted from the ICT images by the specific image processing technology, and then defects in the extracted image could be intelligently recognized by SSD algorithm.

The rest of this paper is organized as follows. The theory of the proposed method is introduced in the section 2. The result is shown in the section 3 and the discussions should be given too. At the last, the conclusion is summarized in the section 4.

\section{Methods}

\subsection{Framework of I-NDT method}

\begin{figure}[htb]
	\centering
	\includegraphics[width=12cm]{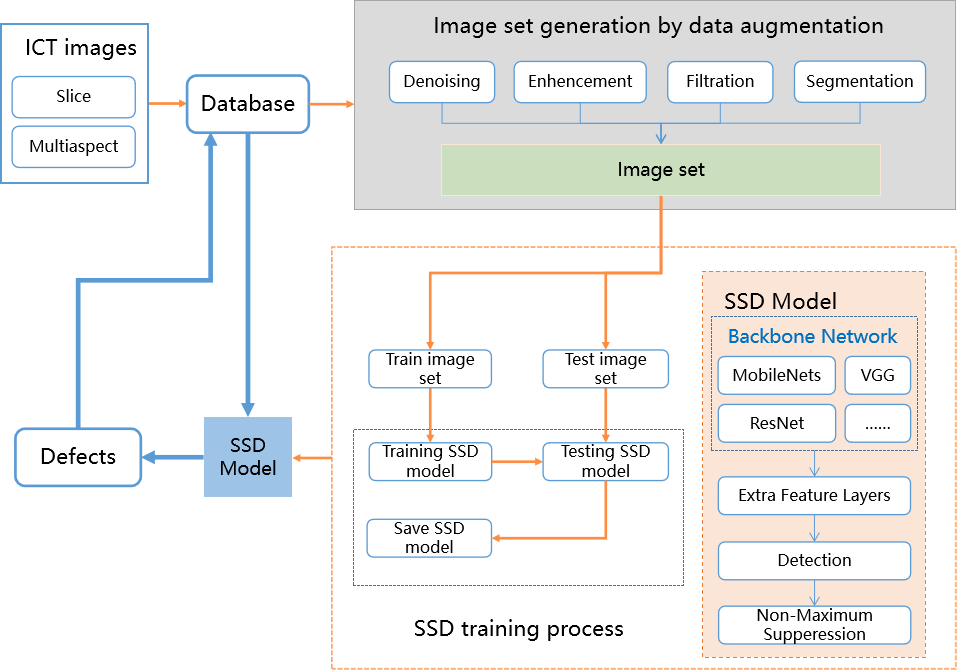}
	\caption{Framework of the I-NDT method}
	\label{fig:framework}
\end{figure}

The I-NDT method is proposed to examine internal defects of power capacitors automatically through ICT. The key point is how to recognize defects from ICT images. Figure\ref{fig:framework} is the framework of the I-NDT method. It is visible that the ICT method contains two main parts: the image set generation by data augmentation and the SSD training process. Obviously, the process of image set generation is used to generate the image set as the input of SSD training process. Once the trained SSD model has satisfied the requirement, the SSD would be used to recognize the defect from the ICT image. Then both the feature of the defect and the ICT image will be saved into a database for further study.

\subsection{Industrial computed tomography image technology}
\begin{figure}[htb]
	\centering
	\includegraphics[width=10cm]{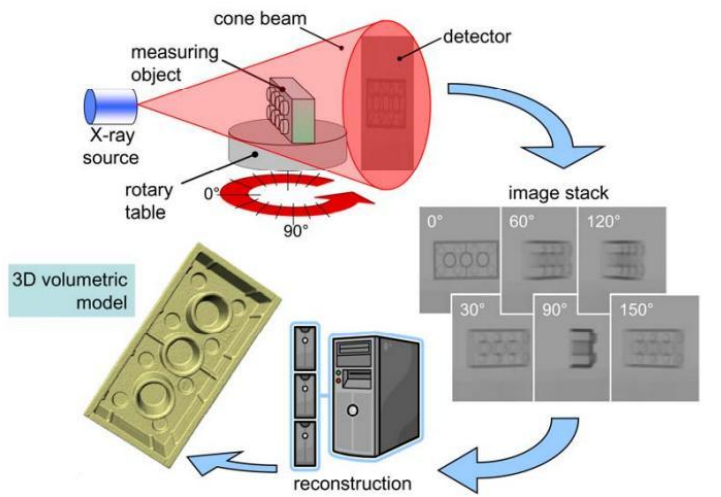}
	\caption{The schematic of ICT reconstruction\cite{sun2012overview}}
	\label{fig:ict}
\end{figure}

Industrial computed tomography image technology is usually used in the detection of material flaws such as voids, cracks and for dimensional measurements. 
ICT systems consist of an X-ray source, a rotary table, an X-ray detector and a powerful data processing unit to 
compute, display and analyse the measurement results. Modern industrial CT systems differ from their medical 
counterparts in so far as the measurement object rotates on a rotary table. The measurement chain of industrial CT 
starts with the X-ray source. Depending on whether a collimated fan beam in combination with a line (1D) detector is used or a cone beam with an area (2D) detector, CT systems measure either 2D information (2D-CT) or 3D information (3D-CT) with one revolution of the part, respectively. To scan an object completely with 2D-CT, a coupled translation of source and detector is necessary as these systems measure the object successively in thin slices. In the case of 3D-CT, objects which fit entirely into the cone beam can be measured with just one revolution of the rotary table. A linear translation of source and detector is not necessary. This is why 3D-CT offers speed advantages compared to 2D-CT. ICT systems generate volumetric images from 2D images of the object under investigation where these image stacks are individually referred to as radiographs (see Fig.\ref{fig:ict}). By reconstructing the image stacks the volumetric image is constructed. A voxel is a unit of graphical information that defines a point in a 3D space. In the case of an ICT scan, a voxel also has a value that represents the density of the material at the point represented by the voxel.

\subsection{Generation of image set based on data augmentation}

Deep neural networks have performed remarkably well on many Computer Vision tasks. However, these networks are heavily reliant on big data to avoid overftting. Overftting refers to the phenomenon when a network learns a 
function with very high variance such as to perfectly model the training data. Unfortunately, many application domains do not have access to big data, such as medical image analysis.

\begin{figure}[htb]
	\centering
	\includegraphics[width=10cm]{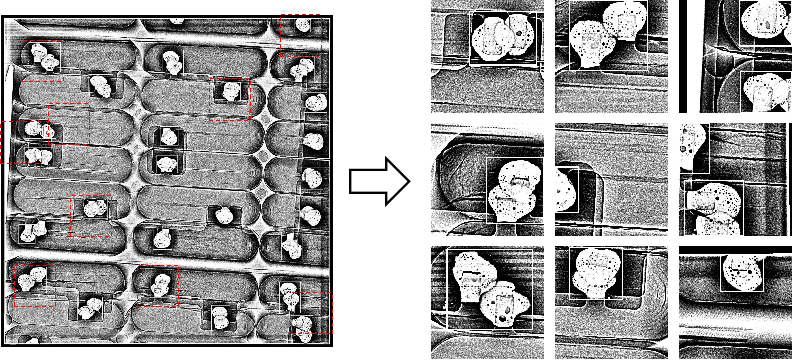}
	\caption{An example of data augmentation for an ICT image}
	\label{fig:data-arg}
\end{figure}

Data augmentation algorithms are used in this study due to the lack of samples. Data Augmentation encompasses a suite of techniques that enhance the size and quality of training datasets such that better Deep Learning models can be built using them. Many image augmentation algorithms have been proposed in recent decades, such as geometric transformations, cropping, rotation, noise injection, color space augmentations, kernel filters, mixing images, neural style transfer, and so on\cite{shorten2019survey}. According to the feature of ICT images of power capacitors, kernel filters and cropping are used in this study. An ICT image will be filtered firstly and then be cropped randomly with size of $512 \times 512$ as shown in Fig.\ref{fig:data-arg}.

\subsection{SSD Architecture}

\begin{figure}[htb]
	\centering
	\includegraphics[width=12cm]{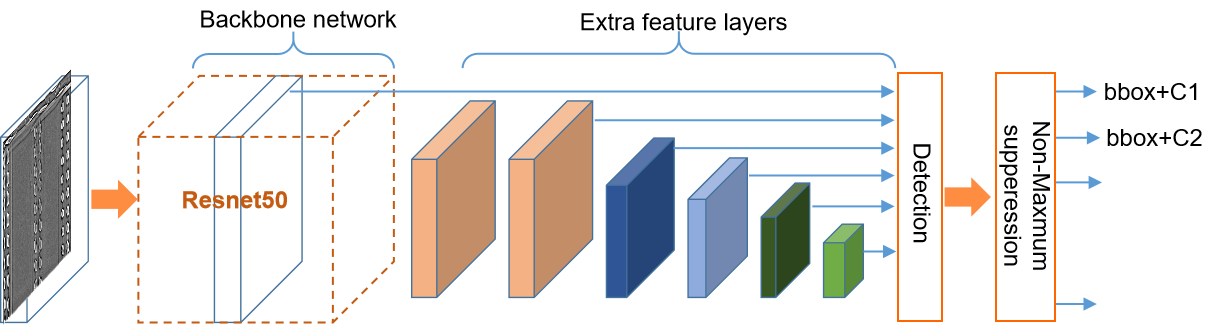}
	\caption{The structure of SSD algorithm}
	\label{fig:ssd}
\end{figure}

The SSD approach is based on a feed-forward convolutional network that produces a fixed-size collection of bounding boxes and scores for the presence of object class instances in those boxes, followed by a non-maximum suppression step to produce the final detection. The SSD is a relatively fast and robust method. It is based on a feed forward convolution network that makes full use of the features of different output layers for object detection. The network structure used for probability of classification (BCS) is shown in Fig.\ref{fig:ssd}. It can be divided into two parts: the backbone network (Resnet50) and the extra feature layers. The front of the network for BCS classification is Resnet50, which has five stages including 49 convolution layers and 1 Fully-Connected (FC) layers. The filters of all layers are used with a very small receptive field: 3 × 3, which is a main contribution to improve the classification ability and decrease the amounts of parameters. The second part of the network is extra feature layers. There are six different scales of feature maps to detect different size of objects. Low-level layers are used to detect small targets, and high-level layers are used to detect targets of large size.

\section{Results and discussions}

\subsection{ICT images of power capacitors}
\begin{figure}[htb]
	\centering
	\includegraphics[width=12cm]{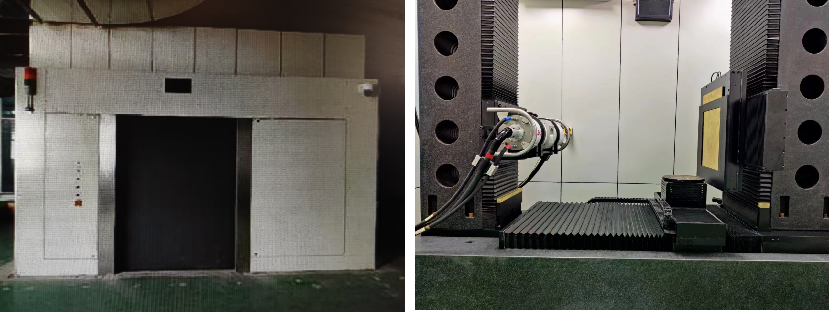}
	\caption{The external view of ICT device used in this study}
	\label{fig:ict-device}
\end{figure}

\begin{figure}[htb]
	\centering
	\includegraphics[width=5cm]{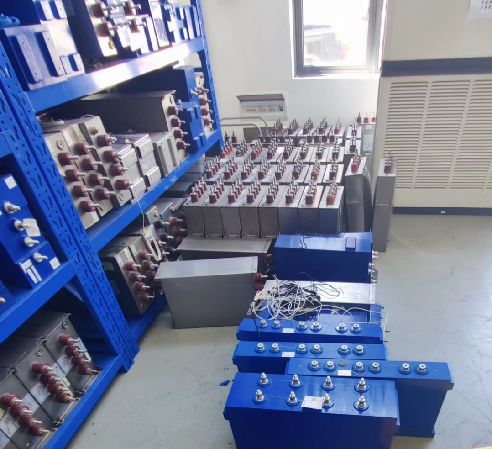}
	\caption{Power capacitors with different type and size}
	\label{fig:pcaps}
\end{figure}

\begin{figure}[htb]
	\centering
	\includegraphics[width=12cm]{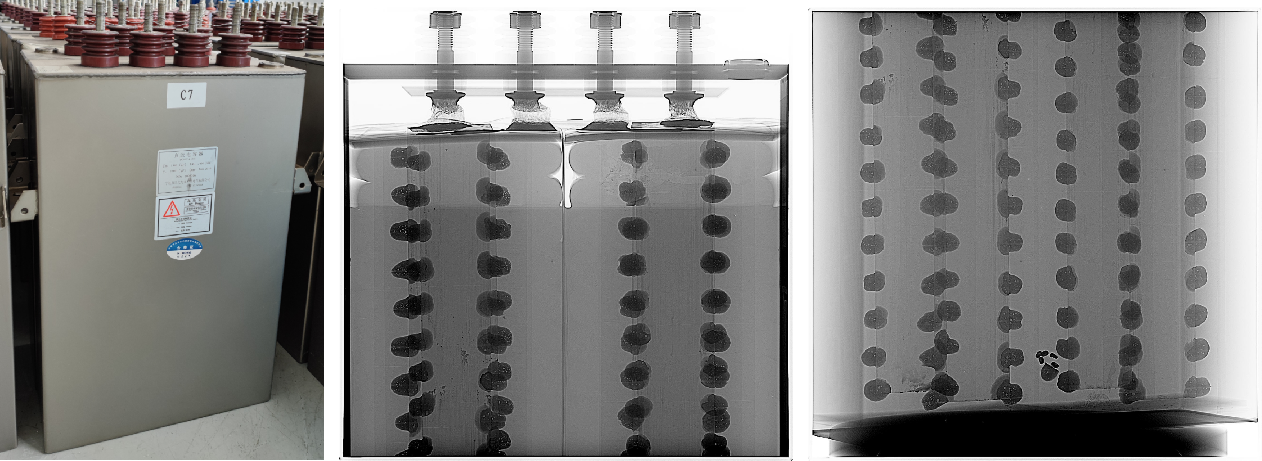}
	\caption{An example of ICT sanning images}
	\label{fig:ict-ex}
\end{figure}

As mentioned above, ICT is used to obtain the internal structure of power capacitors from different aspects. The device used in this study is an industrial X-ray source CT with a 450kV scanner and a fringe projection system as shown in Fig.\ref{fig:ict-device}. The test object is power capacitor and then power capacitors with different type and size (as shown in Fig.\ref{fig:pcaps}) will be tested by ICT one by one. One of the images scanned by ICT is shown in Fig.\ref{fig:ict-ex}. The left picture is the object of power capacitors and the right two pictures are the scanned images processed by the ICT device. In order to show the detail structure of power capacitors especially for welding spots, the size of the output image is set as $4096 \times 4096$ pixels. In this study, 20 power capacitors were scanned by the ICT device, and two images were obtained for each power capacitor at different positions and aspects. Thus, 40 ICT images were obtained but these are certainly not enough for SSD training. Therefore, the data augmentation algorithm is used to extend the image set according to the feature of power capacitors which has many repeat parts inside.Then the image set with 3000 images is obtained finally.

Once the origin image set is obtained, the next step is to label those images for the training process of SSD algorithm. In this study, only 40 images are labeled since all the 3000 images are obtained from those 40 images. A typical ICT image was shown on Fig.\ref{fig:data-arg}. It can be found that the typical defect in the power capacitor is the void in the welding spot. Then those welding spots with voids are labeled and we call them voiding spot in this study.

\subsection{Training results of SSD algorithm}

\begin{figure}[htb]
	\centering
	\includegraphics[width=10cm]{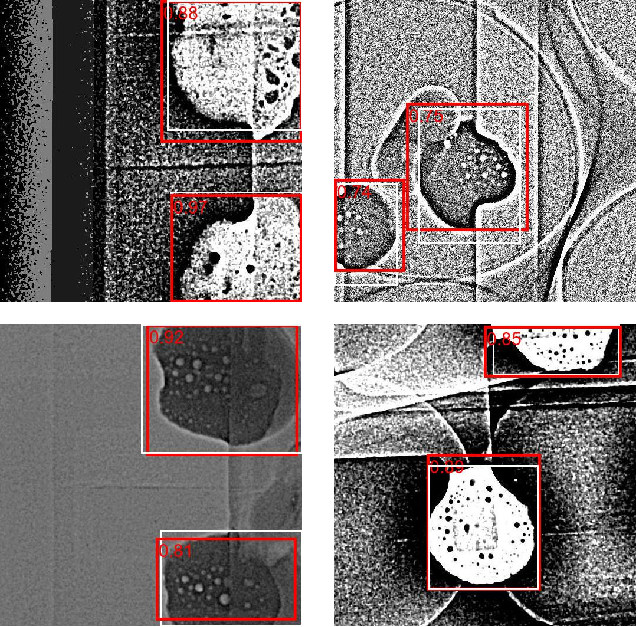}
	\caption{The testing result of the trained model}
	\label{fig:test}
\end{figure}

In this study, $90\%$ of images is used for training and $10\%$ of the images is used for testing, so the training image set contains 2700 images while the testing image set contains 300 images. Then the training and testing image sets are used to train the SSD network where the Resnet50 is used as the backbone network. After the training process completed, the trained SSD model should be tested and then used to predict the defects in a new ICT image. The test result was shown in Fig.\ref{fig:test}, where the ground true was marked with white rectangular while the predicted box was marked with red rectangular. The value of IoU (Intersection over Union) was also shown in the left-up corner of the rectangular. The MIoU (Mean Intersection over Union) is 0.86 over those 300 testing images.

\begin{figure}[htb]
	\centering
	\includegraphics[width=12cm]{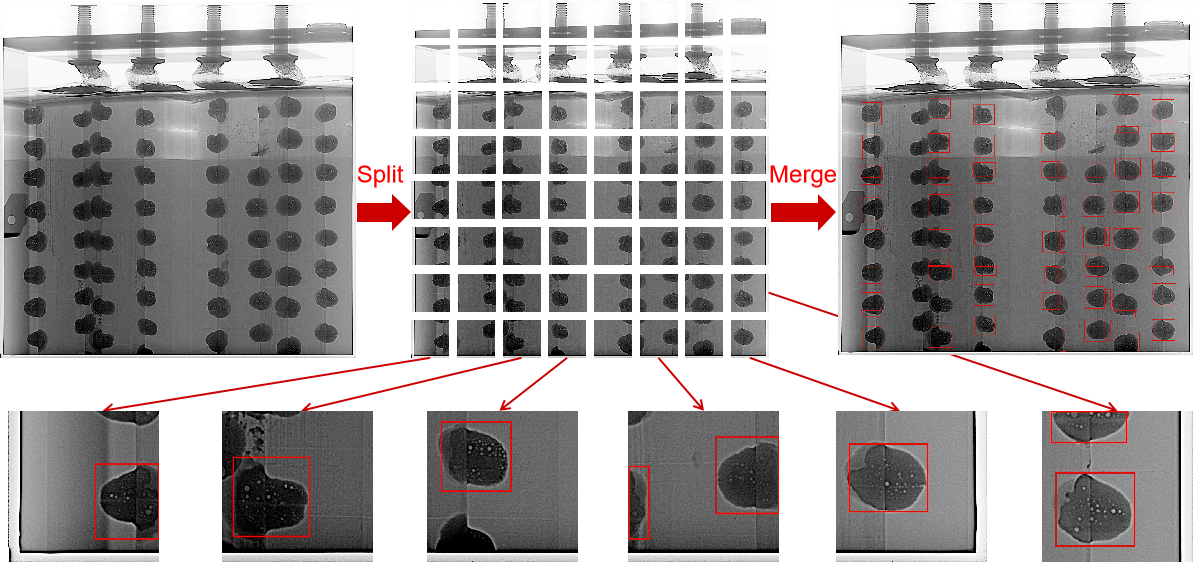}
	\caption{The result of defects recognition by I-NDT method}
	\label{fig:result}
\end{figure}

Moreover, Fig.\ref{fig:result} shows a complete testing process after the SSD model was trained. It can be found that defects (voiding spots) in the powder capacitor are recognized by the I-NDT method and the defects are marked with rectangular in the images. It should be noted that the test ICT image must be separated into $8 \times 8$ images before testing because the feature of welding spots would be ignored in the huge size ($4096\times4096$ pixels) of the ICT image. Then the separated images ($512 \times 512$ pixels) should be tested by the SSD model one by one and defects in the image would be marked with red rectangular.

\section{Conclusion}

In this study, an intelligent non-destructive testing (I-NDT) method based on ICT is proposed to test the quality of power capacitors automatically. The main idea of I-NDT method is to scan the internal structure of power capacitors by the ICT device firstly and then recognize internal defects from ICT images by the SSD algorithm. According to the feature of powder capacitors, 40 images were extended to an image set with 3000 images by data augmentation algorithm to promise the stability and accuracy of the trained SSD model. The test result shows that the propose I-NDT method could recognize the internal defects of power capacitors from the ICT image with acceptable accuracy.

\section*{Acknowledgment}

The authors would like to acknowledge the financial support of the ZhuZhou CRRC Times Electric Co., Ltd. with the project (No.K10RZ22Q1CT0).


\bibliography{mybibfile}

\end{document}